\documentclass[12pt]{article}
\pdfoutput=1
 \usepackage[nottoc]{tocbibind}
\usepackage[colorlinks=true,linkcolor=blue,urlcolor=magenta, citecolor=blue]{hyperref}
\usepackage{fullpage}
\usepackage{amssymb,graphicx}
\usepackage{amsmath}
\usepackage[margin=0.5cm]{caption}
\usepackage{hyperref} 
\usepackage{cite}
\usepackage{upgreek}

\def\e{{\rm e}}

\def\d{\partial}
\def\l{\left(}
\def\r{\right)}

\newcommand{\be}{\begin{equation}}
\newcommand{\ee}{\end{equation}}
\newcommand{\bea}{\begin{eqnarray}}
\newcommand{\eea}{\end{eqnarray}}
\newcommand{\bg}{\begin{gather}}
\newcommand{\eg}{\end{gather}}
\newcommand{\bseq}{\begin{subequations}}
\newcommand{\eseq}{\end{subequations}}

\begin{document}
\baselineskip=15.5pt
\begin{titlepage}
\begin{center}
{\Large\bf  Four Loop Scattering in the Nambu--Goto Theory}\\
%{\Large\bf A Hint of Integrability for the QCD String}\\
%\vspace{0.5cm}
 %The Existence Proof}\\
\vspace{0.5cm}
{ \large
Peter Conkey and Sergei Dubovsky
}\\
\vspace{.45cm}
{\small  \textit{   Center for Cosmology and Particle Physics,\\ Department of Physics,
      New York University\\
      New York, NY, 10003, USA}}\\ 
      \vspace{.1cm}
\end{center}
\begin{center}
\begin{abstract}
We initiate the study of multiloop scattering amplitudes in the Nambu--Goto theory on the worldsheet of a non-critical string. We start  with a brute force calculation of  two loop four particle scattering.  Somewhat surprisingly, even though non-trivial UV counterterms are present at this order,  on-shell amplitudes remain polynomial in the momenta of colliding particles. We show that this can be understood as a consequence of existence of certain close by (semi)integrable models. Furthermore, these arguments can be extended to obtain the answer for three and four loop scattering, bypassing the brute force calculation. The resulting amplitudes  develop non-polynomial (logarithmic) dependence on the momenta starting at three loops.

\end{abstract}
\end{center}
\end{titlepage}
%\tableofcontents
\newpage
\section{Introduction}
Scattering amplitudes provide the most fundamental set of observables
 in a quantum field theory. 
In recent years a dramatic progress has been achieved in developing efficient tools for perturbative multiloop calculations (see, e.g., \cite{Dixon:2011xs,Elvang:2013cua} for an overview) and a number of intriguing insights has been gained building upon the structure of perturbative Feynman diagrams.
These include dual superconformal invariance \cite{Drummond:2008vq} and positive Grassmannian structure \cite{ArkaniHamed:2012nw}
 of $N=4$ supersymmetric Yang--Mills theory, as well as color/kinematics duality \cite{Bern:2010ue} and enhanced ultraviolet cancellations 
 \cite{Bern:2014sna} of supergravity amplitudes. Much of the efforts were focused on highly supersymmetric theories, although the developed techniques have also proven to be very efficient in the studies of hadronic processes as relevant for the LHC \cite{Berger:2008sj}.
 
In this paper we initiate the study of  multiloop amplitudes for yet another theory. As we explain shortly, this study is motivated both by purely theoretical considerations and by the QCD applications. The theory is non-supersymmetric, and unlike the examples mentioned before, it lives in a two-dimensional space-time. This is the world-sheet theory of a bosonic string propagating in a flat $D$-dimensional  space-time. At the leading order in derivative expansion the theory is described by the Nambu--Goto action. In order to set up a scattering problem one picks an infinitely long string as a background, and studies the scattering of small perturbations (``wiggles") propagating along the string. After fixing the static gauge one arrives then at the following action for the physical transverse excitations $X^i$ ($i=1,\dots, D-2$) of a string
 \be
 \label{NGaction}
 S_{NG}=-\ell_s^{-2}\int\;d^2\sigma\sqrt{-\det h_{\alpha\beta}}\;,
 \ee
 where 
 \[
 h_{\alpha\beta}=\eta_{\alpha\beta}+\ell_s^2\d_\alpha X^i\d_\beta X^i
 \]
 is the induced metric on the string worldsheet, and $\ell_s$ is the string width.
 The study of scattering amplitudes in this theory has been initiated in \cite{Dubovsky:2012sh} and motivated by the following question. At any value of $D$ the Nambu--Goto action (\ref{NGaction}) 
 describes a healthy non-renormalizable effective  field theory. The question is how an effective field theorist studying this theory, who is not smart enough to come up with the light cone quantization or Polyakov formalism, will discover that something special happens at the critical number of dimensions, $D=26$?
 
 A (partial) answer to this question can be shortly summarized as follows (for details see \cite{Dubovsky:2012sh,Dubovsky:2012wk,Dubovsky:2015zey}). At tree level the Nambu--Goto theory is integrable 
 ({\it i.e.},  particle production is absent) for any $D$. However, only for a critical string with $D=26$ (and also for $D=3$) can the integrability be preserved at the full quantum level. The corresponding $S$-matrix is entirely determined by a phase shift in two particle scattering, which turns out to be independent of the flavor of colliding particles and takes the following simple form\footnote{Here GGRT stands for Goddard, Goldstone, Rebbi and Thorn \cite{Goddard:1973qh}, because for general $D$ $S$-matrix (\ref{exp}) describes a light cone quantized string.},
 \be
 \label{exp}
 \e^{2i\delta_{GGRT}(s)}=\e^{i\ell_s^2s/4}\;,
 \ee
 where $s$ is the conventional Mandelstam variable. So at $D=3$ and $D=26$ a naively non-renormalizable theory (\ref{NGaction}) gives rise to scattering amplitudes well-defined at all energies. 
 
 As discussed in detail in \cite{Dubovsky:2012wk}, the $S$-matrix (\ref{exp}) exhibits a number of remarkable  properties indicating that it describes a gravitational theory, rather than a conventional field theory. In particular, it exhibits a novel asymptotic behavior at high energies (dubbed ``asymptotic fragility"), which is characterized by the absence of an ultraviolet (UV) conformal fixed point, absence of sharply defined local observables and time delays proportional to the center of mass energy of the collision, in agreement with the Hawking evaporation time in two-dimensional gravity. This raises an intriguing question how to construct non-integrable asymptotically fragile theories. If they exist they may provide a useful laboratory to address the notorious puzzles of black hole physics. This question provides a major motivation for the present brute force study of scattering amplitudes at general $D$.
 
 The question appears very hard by the very nature of asymptotic fragility. Indeed, classification of conventional renormalization group (RG) flows originating from a UV conformal fixed point starts with  specifying all possible relevant deformations of the corresponding UV conformal field theory (CFT).
 This is straightforward for weakly coupled CFTs, and doable 
 (even if challenging)  in many cases for strongly coupled fixed points as well.
  Asymptotically fragile theories do not give rise to a UV CFT, and do not allow to define local operators, so it is not clear a priori whether they can be deformed at all. Still it is tempting to go beyond integrable examples of asymptotic fragility by perturbing around the exact solution (\ref{exp}).
 
To get a sense of the challenge, note that even in the critical case the existence of the UV complete $S$-matrix (\ref{exp}) does not imply that the straightforward perturbative expansion
 in the Nambu--Goto theory is free of the UV divergences.  Instead, after choosing a specific renormalization scheme one expects to encounter an infinite set of higher-dimensional counterterms even at $D=3,26$ (this expectation will be confirmed by explicit calculations presented later in the paper). 
However, these can be fixed by imposing an extra requirement, namely by insisting on integrability.
 
 In other words to construct a theory one needs to find a question to which it provides the answer.
 Note that this situation is not specific at all to the realm of quantum field theories. For instance, a random number on a real axis may be characterized by an infinite sequence of its digits (``counterterm coefficients"). Analogously to a garden variety non-renormalizable theory most numbers cannot be characterized in any sharp way, {\it i.e.} they do not provide an answer to any meaningful question\footnote{Clearly, to qualify as meaningful, a question should at least have a finite length. Hence
  there is  at most a countable set of meaningful questions versus a continuum of numbers.}.
  Some numbers, such as rational ones, provide answers to rather straightforward and boring questions.
  Other numbers, such as $\pi$, $e$, or $\zeta(3)$, provide answers to more sophisticated and interesting questions.
   
 To arrive at the $S$-matrix (\ref{exp}) one may ask 
  ``Are there integrable theories of $D-2$ massless bosons with non-linearly realized $ISO(1,D-1)$ symmetry?". To construct non-integrable asymptotically  fragile theories one needs to look for a less restrictive version of this question, which still would allow fixing (almost) all of the counterterms.
 At the moment we do not know what this new question is, so one reason to take a careful look at the properties of the perturbative diagrams is that they provide  ``experimental" data, which might help to guess the correct question. 
 We would like to give up integrability, however it is natural to try to keep the other conditions, which explains why we decided to take a look at the non-critical Nambu--Goto.
  
 It is worth noting that an unusual and interesting property of the $S$-matrix (\ref{exp}) is that its perturbative expansion in $\ell_s^2$ absolutely converges at all values of energy. In other words, this $S$-matrix is literally given by a sum of the corresponding Feynman graphs, without any non-perturbative effects. This serves as an additional motivation to take a look at the perturbative amplitudes in a non-critical case; it would be remarkable if there were non-integrable theories with this property.
 
 To conclude with motivations, let us stress that  
 an independent reason to develop computational techniques allowing to construct and analyze non-integrable asymptotically fragile theories comes from QCD. Understanding the planar limit of confining theories, such as pure gluodynamics, is a longstanding and fascinating problem. One aspect of this problem is to find a description of the worldsheet theory of confining strings. 
 Time delay corresponding to (\ref{exp}) has a very transparent geometrical origin---a physical length of a relativistic string is proportional to its length.
 A worldsheet theory of a confining string preserves two-dimensional unitarity up to arbitrarily high energies in the planar limit, and one expects it to exhibit  time delays growing with energy.
 Generically, one does not expect a planar worldsheet theory to be integrable.
  Hence it is very plausible that confining strings in general give rise to non-integrable asymptotically fragile theories in the planar limit. These theories may be studied ``experimentally" using lattice simulations \cite{Athenodorou:2010cs,Athenodorou:2011rx,Athenodorou:2013ioa,Athenodorou:2016kpd}.  A recently
 developed TBA approach \cite{Dubovsky:2013gi,Dubovsky:2014fma}  allows to relate
 this data to the worldsheet $S$-matrix. This allows to reconstruct the worldsheet action at small and intermediate energies $s\ell_s^2\lesssim 1$. However, to extend this success to higher energies, it will be useful to develop tools allowing to work with asymptotically fragile theories  in the UV regime  $s\ell_s^2\gg 1$.
 
 The rest of the paper is organized as  follows. In section~\ref{sec:prior} we 
 set up the notations, discuss some generalities about how the calculation is organized and
 review the previously available tree level and one loop results. At this order no non-vanishing on-shell 
 counterterm is present, apart from the evanescent Einstein term, which is a total derivative  for a physical number of worldsheet dimensions, $d=2$.
 
 In section~\ref{sec:2loops} we present a brute force two loop calculation of the two-to-two scattering amplitude. This calculation is instructive from the technical viewpoint, since it  explicitly illustrates the role of the evanescent counterterm which was found at one loop (similar observations have been recently made for gravitational amplitudes in \cite{Bern:2015xsa}). At this order two non-trivial counterterms are allowed, and need to be introduced to obtain a finite answer for the amplitude. However, the amplitude has a surprising property that  the corresponding coupling constants  
  do not run, {\it i.e.} no $\log\mu$ dependence arises in the finite part of the amplitude despite the presence of ${1/ \epsilon}$ poles (we always work in dimensional regularization to preserve the non-linearly realized Poincar\'e symmetry of the Nambu--Goto action).
 
 We are not aware of any symmetry reason for this surprising cancellation. However, in section~\ref{sec:4loops} we manage to explain it as a consequence of the validity of the gravitational dressing procedure introduced in \cite{Dubovsky:2013ira}. In fact, our argument provides a very efficient shortcut to reproduce the whole two loop amplitude without performing the actual calculation.  Furthermore, a straightforward extension of the argument allows to predict the amplitude up to four loops  as a result of one-loop calculation. We find that the cancellations observed at two loops do not persist at higher orders. New counterterms arising at three loop level exhibit a non-trivial $\log\mu$ dependence.
We discuss future directions in section~\ref{sec:last}.
\section{Generalities and a review of prior results}
\label{sec:prior}
\begin{figure}[t!!]
	\centering
	\includegraphics[width=3in]{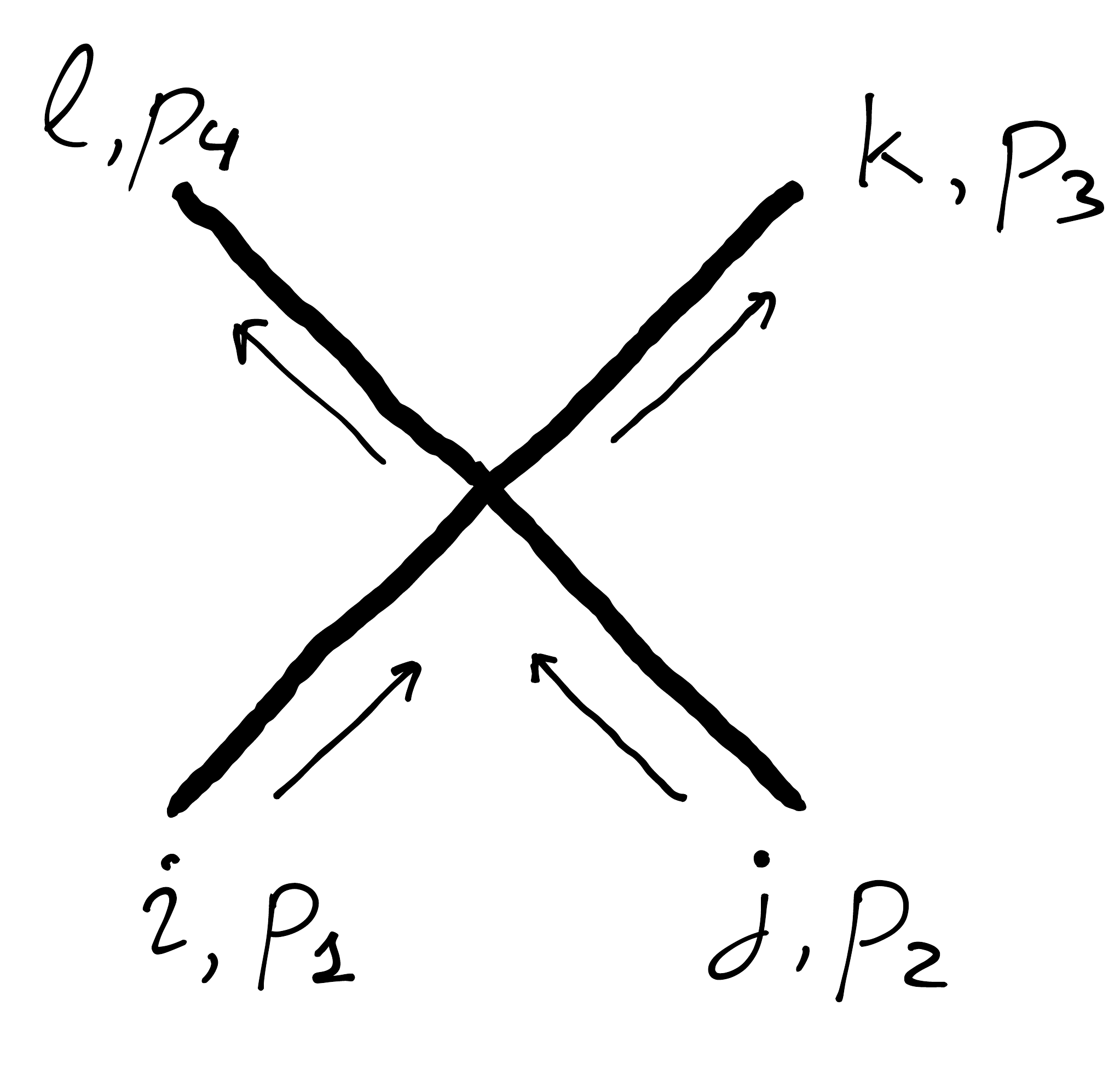}
	%Figures/tree.pdf}
	\caption{Kinematics considered in the paper.}
		\label{fig:kinematics}
\end{figure}

Let us start with several general comments about our conventions and on how the calculation is organized. We are using dimensional regularization throughout this paper. This is the only regularization we are aware of, which preserves the nonlinearly realized Poincare symmetry of the Nambu--Goto action,
\be
\label{deltaX}
\delta^{\alpha i}X^j=-\epsilon\l\delta^{ij}\sigma^\alpha+X^i\d^\alpha X^j\r\;.
\ee
Given that an asymptotically fragile theory is not expected to behave any different from a garden variety non-renormalizable effective theory as far as off-shell observables are concerned, we are only interested in on-shell renormalization of the non-critical Nambu--Goto theory. In this paper we restrict to the  two-to-two scattering amplitude, which can be parametrized in the following way,
\be
{\cal M}_{ij,kl}=A\delta_{ij}\delta_{kl}+B\delta_{ik}\delta_{jl}+C\delta_{il}\delta_{jk}\;,
\ee
where $i,j$ ($k,l$) are the flavor labels of incoming (outgoing) particles. The corresponding momenta are $p_1,p_2$ and $p_3,p_4$, so that the Mandelstam invariants are 
\[
s=-(p_1+p_2)^2\;,\;\;\; t=-(p_1-p_3)^2\;,\;\;\;u=-(p_1-p_4)^2\;.
\]
Without a loss of generality, in what follows we assume that  $p_1,\,p_3$ are right-moving momenta and $p_2,\,p_4$ are left-moving (see Figure~\ref{fig:kinematics}).
Then $t=0$ and $u=-s$ so that the $A,B,C$ amplitudes become functions of a single Mandelstam variable $s$. With this choice of kinematics $A$ gives the annihilation amplitude, $B$ corresponds to transmission and $C$ to reflection.

Note that particle production is absent in the Nambu--Goto theory at the tree level, so the first inelastic process (two-to-four scattering) arises only at ${\cal O}(\ell_s^6)$ order. The corresponding amplitude has been calculated in  \cite{Cooper:2014noa}. As a result 
the two-to-two unitarity is first broken only at the ${\cal O}(\ell_s^{12})$ (five loop) order and provides a useful guidance at lower orders.  In terms of functions $A$, $B$ and $C$ the two particle unitarity conditions read as follows
\begin{gather}
\label{unitarityA}
\mbox{Im} A={1\over 4s}\l2|A|^2+AB^*+A^*B+AC^*+A^*C\r\\
\label{unitarityB}
\mbox{Im} B={1\over 4s}\l|B|^2+|C|^2\r\\
\mbox{Im} C={1\over 4s}\l B^*C+BC^*\r\;.
\label{unitarityC}
\end{gather}

A somewhat unusual property of massless scattering, which is useful to keep in mind, is that these amplitudes have a cut starting at the origin and extending over all real axis, which makes it impossible to perform an analytic continuation to the lower half of the physical sheet in the complex $s$-plane without passing through a singularity. Of course, one can still analytically continue the $A,B,C$ functions into the $\mbox{Im}\,s<0$ region, however, after this continuation the lower half of the $s$-plane is not a part of the physical region.
 Nevertheless, crossing symmetry and real analyticity requirements still impose useful constraints on these amplitudes, which read as
 \begin{gather}
 \label{crossingA}
 C(s)=A(\e^{i\pi}s)^*\\
 B(s)=B(\e^{i\pi}s)^*\;,
 \label{crossingB}
 \end{gather}
 where $A(\e^{i\pi}s)$ stands for the analytic continuation from $s$ to $-s$ through the upper half plane.
 
%Amplitudes $A$, $B$ and $C$ are related by crossing symmetry as 
%\be
%\label{crossing}
%A(s,t,u)=A(s,u,t)=B(t,s,u)=C(u,t,s)\;,
%\ee
%and in what follows we will be presenting results for the $A$ (annihilation) amplitude only. As a consequence of two dimensional kinematics one has either $t=0$ or $u=0$. However, we will keep terms proportional to $t u$ because
%they give rise to non-vanishing contributions in other channels. On the other hand, the $stu$ product vanishes in all channels and we will drop terms proportional to $stu$, unless stated otherwise.

Previously, tree level and one loop amplitudes were calculated in \cite{Dubovsky:2012sh}. At tree level one finds that only the transmission amplitude is non-zero and it is given by
\be
\label{tree}
B_2={\ell_s^2\over2}s^2\;.
\ee
At one loop one finds two physically distinct contributions into the amplitude. First, there is a rational term, 
which corresponds to the Polchisnki--Strominger (PS) interaction \cite{Polchinski:1991ax}, contributing to annihilations and reflections,
\be
\label{PS1loop}
A_4=-C_4=-{D-26\over 192\pi}\ell_s^4s^3\;.
\ee
Second, as required by unitarity, the transmission amplitude acquires an imaginary part,
\[
B_4=i{\ell_s^4\over 16}s^3\;.
\]
The annihilation/reflection part obviously vanishes at $D=26$. It also  vanishes in a single flavor ($D=3$) case, because  the only physical amplitude is then given by the sum $A+B+C$.
Both for $D=26$ and $D=3$ the remaining transmission amplitude agrees with the phase shift (\ref{exp}), given that after accounting for the proper normalization of states the two are related as (see \cite{Dubovsky:2012wk} for the details on all factors of 2)
\[
\e^{2i\delta}=1+{i B\over 2 s}\;.
\]
Note that all physical on-shell amplitudes are finite at one-loop. This follows from the absence of non-trivial ${\cal{O}}(\ell_s^4)$ counterterms compatible with non-linearly realized Poincar\'e symmetry.
However, if one keeps external momenta in $d=2-2\epsilon$ dimensions ({\it i.e.}, does not impose the two dimensional relation $stu=0$), one finds the following divergent contribution into the amplitude,
\be
\label{Einsteininf}
M_{ij,kl}\supset -{(D-8)\ell_s^4\over96\pi \epsilon}stu\l\delta_{ij}\delta_{kl}+\delta_{ik}\delta_{jl}+\delta_{il}\delta_{jk}\r\;.
\ee
This contribution is related to the presence of an evanescent operator, the Einstein--Hilbert term, which turns into a total derivative at $d=2$.
To cancel the divergence (\ref{Einsteininf}) one needs to add the following counterterm to the action,
\be
\label{Eins}
S_{E_1}=-{C_{E_1}\mu^{2\epsilon}\over 48\pi\epsilon}\int d^{d}\sigma\sqrt{-h}R={C_{E_1}\mu^{2\epsilon}\ell_s^4\over 48\pi\epsilon}\int d^{d}\sigma(\d_\beta X^i\d_\alpha\d_\gamma X^i)(\d^\gamma X^i\d^\alpha\d^\beta X^i)+\dots\;,
\ee
where 
\[
C_{E_1}=D-8\;.
\]
However, from the viewpoint of calculating the physical on-shell amplitudes, which is our goal here, nothing forces us to include  this counterterm yet. So it will be instructive to see that the same counterterm is also required for the consistency of the physical on-shell two loop amplitudes. 

\section{Brute force two loop calculation}
\label{sec:2loops}
\begin{figure}[t!!]
	\centering
	\includegraphics[width=5in]{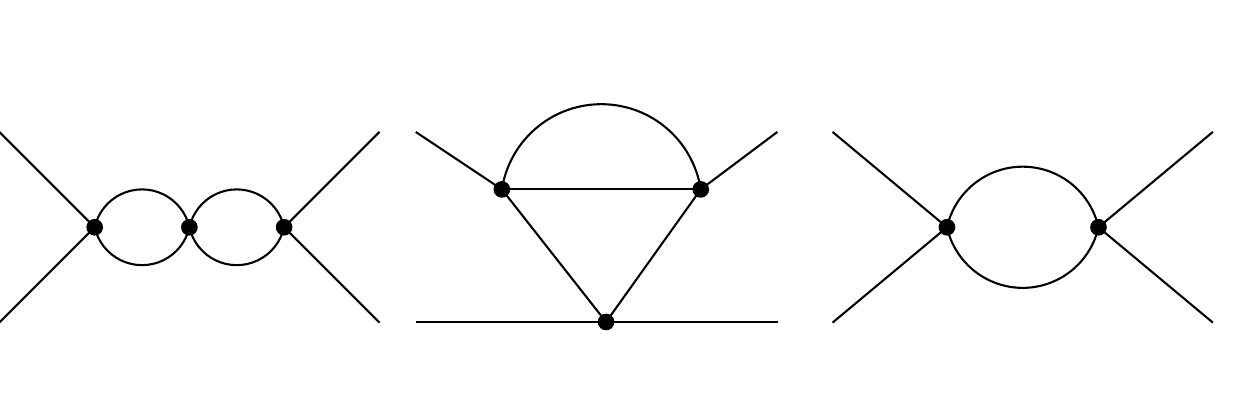}
	\caption{Topologies contributing at order $l_s^6$.}
		\label{fig:topologies}
\end{figure}

Let us describe now the two loop calculation and its results. The calculation itself is rather straightforward, even if somewhat tedious. It was carried out in Mathematica  with assistance from FeynRules \cite{FeynRules},
FeynArts \cite{FeynArts} and FeynCalc \cite{FeynCalc}. There are three different topologies contributing at this order, as illustrated in Figure~\ref{fig:topologies}. These are two loop double fishes and wine glass diagrams, and also one loop fish diagrams, where one of the vertices  corresponds to the evanescent counterterm (\ref{Eins}). 

At this order one finds two non-trivial counterterms in the Lagrangian,
\begin{gather}
S_1=C_1\mu^{2\epsilon}\ell_s^6\int d^d\sigma\sqrt{-h}K^i_{\alpha\beta}K^i_{\delta\gamma}K^{j\alpha\beta}K^{j\delta\gamma}
\nonumber
\\
=C_1\mu^{2\epsilon}\ell_s^6\int d^d\sigma(\d_\alpha\d_\beta X^i\d_\delta\d_\gamma X^i)(\d^\alpha\d^\beta X^j
\d^\delta\d^\gamma X^j)+\dots
\label{S1}
\end{gather}
and 
\be
S_2=C_2\mu^{2\epsilon}\ell_s^6\int d^d\sigma\sqrt{-h}\l K^i_{\alpha\beta}K^{i\alpha\beta}\r^2
%K^j_{\delta\gamma}K^{j\delta\gamma}
=C_2\mu^{2\epsilon}\ell_s^6\int d^d\sigma(\d_\alpha\d_\beta X^i\d^\alpha\d^\beta X^i)^2
%(\d_\delta\d_\gamma X^j\d^\delta\d^\gamma X^j)
+\dots\;,
\label{S2}
\ee
where $K^i_{\alpha\beta}$ is the extrinsic curvature of the worldsheet.
Hence one expects to find UV divergences in physical on-shell amplitudes at this order. Indeed, one finds 
\begin{gather}
A_6^\infty=C_6^\infty=-{(D-8-2C_{E_1})(D-12)\ell_s^6\over 9216\pi^2\epsilon }s^4\\
B_6^\infty={(D-8-2C_{E_1})\ell_s^6\over 768\pi^2\epsilon }s^4\;,
\end{gather}
which can be canceled with the choice
\be
\label{C12}
C_1={2C_{E_1}-D+8\over 384\pi^2\epsilon}\;,\;\;C_2={(D-6)(D-8-2C_{E_1})\over 4608\pi^2\epsilon}\;.
\ee
Explicit expressions for finite parts of amplitudes for a general value of $C_{E_1}$ are too ugly to provide them here, so we will present only their physical values at  \[ C_{E_1}=D-8\;.\]
However, it is worth pointing out that, as anticipated at the end of Section~\ref{sec:prior}, this value of $C_{E_1}$ indeed does get fixed from the properties of the physical two loop amplitudes. Namely, for any other value of $C_{E_1}$
the unitarity conditions (\ref{unitarityA}), (\ref{unitarityB}), (\ref{unitarityC})  fail to be satisfied. Note that the physical value of $C_{E_1}$ is different from $C_{E_1}=(D-8)/2$ which would set to zero the UV counterterms (\ref{C12}), which is not surprising. It is somewhat surprising though that all UV divergences encountered so far vanish at $D=8$, but this appears to be a coincidence. 

For the physical value of $C_{E_1}$ finite parts of the on-shell two loop amplitudes take the following form
\begin{gather}
\label{A6}
A_6(s)=C_6(-s)^*=- {D-26\over 768 \pi}i\ell_s^6s^4-{6D^2-143D+448\over 13824 \pi^2}\ell_s^6s^4\\
\label{B6}
B_6(s)=-{\ell_s^6s^4\over 192}+11{D+4\over 4608\pi^2}\ell_s^6s^4\;.
\end{gather}
The most surprising property of these amplitudes is that they are polynomial in $s$. Given the presence of two non-trivial UV counterterms one would expect the amplitudes to depend on the renormalization scale $\mu$, and as a consequence, also on $\log s$. This did not happen.  It was noticed recently for gravitational amplitudes \cite{Bern:2015xsa}, that in the presence of evanescent divergences the coefficient in front of the leading UV $1/\epsilon$-pole is in general different from the coefficient in front of the leading $\log\mu$ dependence. We also observe it here, but a really peculiar property of the Nambu-Goto amplitudes is that $\log\mu$ dependence automatically cancels out at two loop order. 

We are not aware of any conventional symmetry argument for this cancellation. However, as we argue in the next section, it can be explained based on the existence of certain close by (semi)integrable models. Furthermore these arguments will allow us to reconstruct without much calculations
not only the two loop  amplitude, but also three and four loop amplitudes.

Finally, let us point out that before imposing the two-dimensional kinematical restriction $stu=0$, on-shell two loop amplitudes contain additional divergence of the form
\be
\label{2loopevan}
M_{ij,kl}\supset{\ell_s^6\over 4608\pi^2}\l{D(D-8)\over\epsilon^2}-{8(7D-62)\over\epsilon}\r stu \l s\delta_{ij}\delta_{kl}+t\delta_{ik}\delta_{jl}+u\delta_{il}\delta_{jk}\r\;.
\ee
Similar to what we found at one loop, this indicates the presence of a new evanescent operator, and this operator will be required to ensure unitarity of physical amplitudes at higher loop orders.
This new evanescent operator is related to the identity 
\be
G_{\alpha\beta}\equiv R_{\alpha\beta}-{1\over 2}g_{\alpha\beta}R=0\;,
\ee
which holds in two dimensions. This identity implies that any operator proportional to the Einstein tensor $G_{\alpha\beta}$ is evanescent. In particular, at two loop order we find the following evanescent operator
\begin{gather}
S_{E_2}=C_{E_2}\mu^{2\epsilon}\int d^d\sigma R^{\alpha\beta}G_{\alpha\beta}
\nonumber
\\
=C_{E_2}\mu^{2\epsilon}\int d^d\sigma\l-{1\over 2}\l\d_\alpha\d_\beta X^i\d^\alpha\d^\beta X^i\r^2
+\d_\gamma\d_\alpha X^i\d^\gamma\d_\beta X^i\d_\delta\d^\alpha X^j\d^\delta\d^\beta X^j\r\;.
\end{gather}
Requiring cancellation of the evanescent divergence (\ref{2loopevan}) fixes
\be
C_{E_2}={D(D-8)\over2304\pi^2\epsilon^2}+{196-23 D\over 1728\pi^2\epsilon}\;.
\ee 

\section{Four loops  from gravitational (un)dressing}
\label{sec:4loops}
The surprising feature of the physical amplitudes (\ref{A6}) and (\ref{B6}) is that they take polynomial form, {\it i.e.} no $\log\l s/\mu^2\r$ dependence arises despite the presence of physical UV divergences.
We are not aware of any conventional symmetry explanation for this. Nevertheless, as we will see now,
this property can be understood as a consequence of the existence of certain (semi)integrable
theories. 

Let us first present the argument for the critical string case $D=26$ (which also applies at $D=3$), where it applies in the most straightforward way. As we explained in the {\it Introduction}, for these values of $D$ the phase shift (\ref{exp}) defines an integrable theory enjoying non-linearly realized target space Poincar\'e symmetry. This implies that one should be able to reproduce the corresponding $S$-matrix perturbatively, starting with the Nambu--Goto action and appropriately choosing the finite part of coefficients for higher order counterterms. No Poincar\'e invariant counterterm is present at one loop level, so the only freedom at the level of two loop amplitudes is related to counterterms (\ref{S1}), (\ref{S2}), which may only affect the polynomial part of the two loop amplitudes.
Hence, the presence of non-polynomial terms would prevent us from reproducing the $S$-matrix perturbatively from a local Lagrangian, which explains why non-polynomial terms did not get generated at $D=3, 26$.

This argument cannot be straightforwardly applied for other values of $D$. Indeed, in this case to reproduce the integrable $S$-matrix (\ref{exp})  from the local Lagrangian one needs to introduce a one loop counterterm canceling the Polchinski--Strominger amplitude,
\[
S_{PS4}=\ell_s^4{D-26\over192\pi}\int d^2\sigma\d_\alpha\d_\beta X^i\d^\alpha\d^\beta X^i\d_\gamma X^j\d^\gamma X^j\;,
\]
which breaks the non-linearly realized Poincar\'e symmetry. A priori, one would expect that insertion of this counterterm into one loop diagrams should introduce non-polynomial terms in the difference between two loop amplitudes of the Poincar\'e invariant Nambu--Goto theory and of the GGRT theory (\ref{exp}).

The explanation for why this does not happen is related to the existence of the gravitational dressing procedure introduced in 
\cite{Dubovsky:2013ira}.  Gravitational dressing works as follows. One starts with an {\it arbitrary} relativistic two dimensional quantum field theory, characterized by the $S$-matrix elements $S(p_i)$. Here, unlike in the rest of this paper, all the momenta are taken as incoming. Then a gravitationally dressed $S$-matrix is defined by
\be
\hat{S}(p_i)=\e^{\ell_s^2/4\sum_{i<j}p_i*p_j}S(p_i)\;,
\ee
where the $*$-product is
\[
p_i*p_j=\epsilon_{\alpha\beta}p_i^\alpha p_j^\beta\;,
\]
and we made use of the existence of a natural cyclic order (by rapidities) for non-zero two-dimensional momenta. As argued in \cite{Dubovsky:2013ira} the dressed amplitudes described by $\hat{S}$ exhibit all the properties expected from a
 healthy relativistic  $S$-matrix. For example, starting with a theory of $D-2$ free massless bosons  gravitational dressing results in the GGRT $S$-matrix (\ref{exp}).
 
 Now to run the argument let us start with a theory described by the following action,
 \[
 S_0=S_{free}-S_{PS4}\;,
 \]
 where $S_{free}$ is a free kinetic term for $D-2$ massless bosons.  Then, by construction, the corresponding dressed theory $\hat{S}_0$ has the same ${\cal O}(\ell_s^2)$ and ${\cal O}(\ell_s^4)$ four particles amplitudes as the Nambu--Goto theory $S_{NG}$.
 Moreover, at the ${\cal O}(\ell_s^6)$ level the only difference between four particle $\hat{S}_0$ amplitudes and the Nambu--Goto ones may be due to a different choice of $S_1$, $S_2$ counterterms\footnote{Note that even in the absence of non-linearly realized Poincar\'e symmetry $S_1$ and $S_2$ are the only non-trivial quartic $O(D-2)$ invariant ${\cal O}(\ell_s^6)$ counterterms.}, {\it i.e.}, it takes a polynomial form.
 Hence the presence of non-local $\log\l s/\mu^2\r$ terms in the Nambu--Goto amplitudes would prevent one from writing a local Lagrangian matching  $\hat{S}_0$ amplitudes. On the other hand, such a Lagrangian should exist, because $S$-matrix $\hat{S}_0$ exhibits all the required analytic properties. This completes the argument and explains the absence of $\log\l s/\mu^2\r$ contributions in the amplitudes calculated in the previous section.
 
Moreover, this argument immediately allows us to write down the answer for two particle Nambu--Goto amplitudes up to ${\cal O}(\ell_s^6)$ order. Up to non-universal tree level $S_1$ and $S_2$ contributions they should be the same as in the $\hat{S}_0$ theory,
{\it i.e.} should agree with the ${\cal O}(\ell_s^6)$ expansion of 
\begin{gather}
\label{dressedtreeA}
\hat{A}=\e^{i\ell_s^2s/4}A_{PS4}=-{D-26\over 192\pi}\ell_s^4s^3-i{D-26\over 768\pi}\ell_s^6s^4+{\cal O}(\ell_s^8)\\
\label{dressedtreeB}
\hat{B}=\e^{i\ell_s^2s/4}B_{PS4}+2s i\l1-\e^{i\ell_s^2s/4}\r={\ell_s^2s^2\over 2}+i{\ell_s^4s^3\over 16}-{\ell_s^6 s^4\over 192}+{\cal O}(\ell_s^8)\\
\label{dressedtreeC}
\hat{C}=\e^{i\ell_s^2s/4}C_{PS4}={D-26\over 192\pi}\ell_s^4s^3+i{D-26\over 768\pi}\ell_s^6s^4+{\cal O}(\ell_s^8)\;,
\end{gather}
%
%\begin{gather}
%\label{dressedtreeA}
%\hat{A}=-\hat{C}=-{D-26\over 192\pi}\e^{i\ell_s^2s/4}\ell_s^4s^3=-{D-26\over 192\pi}\ell_s^4s^3-i{D-26\over 768\pi}\ell_s^6s^4+{\cal O}(\ell_s^8)\\
%\label{dressedtreeB}
%\hat{B}=2s i\l1-\e^{i\ell_s^2s/4}\r={\ell_s^2s^2\over 2}+i{\ell_s^4s^3\over 16}-{\ell_s^6 s^4\over 192}+{\cal O}(\ell_s^8)\;.
%\end{gather}
where $A_{PS4}$ and $B_{PS4}$ are  ${\cal O}(\ell_s^4)$ amplitudes in the $S_0$ theory.
As expected, these expressions agree with our brute force two loop results (\ref{A6}), (\ref{B6}) up to non-universal polynomial pieces. Of course, given we already established that there can be no non-local real parts, the matching of the imaginary parts 
is also guaranteed by unitarity. It is interesting though that the dressing argument allows to correctly predict also the leading transcendental part of (\ref{B6}) (the term without $\pi^2$ in the denominator). 

Furthermore, the dressing argument can actually be pushed quite a bit further by reversing the logic above in the following way.
 First, gravitational dressing implies the existence of an ``undressed" theory with a local action $S_{PS}$, such that its dressing gives the Nambu--Goto amplitudes (for any choice of higher order counterterms). This follows from the possibility to perform dressing with negative tension
 $\ell_s^2\to -\ell_s^2$ (we will call this undressing).
 The action of the undressed theory can be written as 
 \be
\label{SPS}
S_{PS}=S_{free}-S_{PS4}-S_{PS6}-\dots\;,
\ee
where the ${\cal O}(\ell_s^6)$ sextic vertex  $S_{PS6}$  was calculated in \cite{Cooper:2014noa}. Here dots stand for terms with larger number of fields and derivatives.
 
 For an arbitrary term in the action it is convenient to define its weight $h$
to be equal to the difference between the number of derivatives and the number of fields. The weight is equal to twice the number of loops at which the corresponding counterterm may get generated starting with the tree level Nambu--Goto action (or any other $h=0$ classical action).

For example, $S_{free}$ has $h=0$, $S_{PS4}$ and $S_{PS6}$ interactions in (\ref{SPS}) have $h=2$. By construction 
$S_{PS}$ does not contain $h=1$ terms and all its $h=2$ terms are uniquely fixed by universal one loop Nambu--Goto amplitudes.
Terms with higher weight in $S_{PS}$ are not uniquely fixed, which corresponds to a freedom of choosing different coefficients for higher order counterterms in the Nambu--Goto theory.

 In particular, all $h>2$ quartic vertices in $S_{PS}$ are completely free. Indeed,
  it is straightforward to check that in two dimensions all shift invariant quartic interactions with $h>2$ can be written in such a way that there are at least two derivatives acting on each $X$. Given that the extrinsic curvature of the string worldsheet takes the form
\[
K^i_{\alpha\beta}=\d_\alpha\d_\beta X^i+\dots,
\]
any such term can be written in a form compatible with the non-linearly realized Poincar\'e symmetry \cite{Dubovsky:2015zey}.
Hence the number of free coefficients for $h>2$ quartic vertices in the Nambu--Goto theory exactly matches the number of corresponding free coefficients in the undressed
$S_{PS}$ theory. Hence any choice of $h>2$ quartic vertices in $S_{PS}$ is consistent with the Poincar\'e symmetry of $\hat{S}_{PS}$.

%
%Consequently, the leading order Poincar\'e breaking term which may appear in $\hat{S}_{PS}$ is a $h=4$ sextic vertex. A related Poincar\'e breaking contribution to the four-point amplitude may arise from the three loop diagram, %presented in Figure~\ref{fig:2sextics}, where the second sextic vertex is from the Nambu--Goto part of the action.
%

For example, we may set all these coefficients to zero at weight $h=4,6$. With this choice  the four particle amplitude in $\hat{S}_{PS}$  up to (and including) ${\cal O}(\ell_s^{10})$ order is the same as in $\hat{S}_{0}$ theory. 
It is straightforward to calculate these contributions.
 At ${\cal O}(\ell_s^8)$ order four particle amplitudes in the $S_{0}$ theory are given by one loop fish diagrams with $S_{PS4}$ vertices. Evaluating these diagrams gives the following result,
 \begin{gather}
 \label{APS8}
 A_{PS8}={(D-26)^2\over 4\pi (192 \pi)^2}\ell_s^8s^5\l\l D-4\r ( i\pi-\log{s\over\mu^2})-{89\over 30}\r \\
 \label{BPS8}
 B_{PS8}=i{(D-26)^2\over 4(192 \pi)^2}\ell_s^8s^5\\
 \label{CPS8}
 C_{PS8}={(D-26)^2\over 4\pi (192 \pi)^2}\ell_s^8s^5\l \l D-4\r\log{s\over\mu^2}+{89\over 30}\r\;.
 \end{gather}
 The corresponding Poincar\'e invariant amplitudes can be obtained by adding (\ref{APS8}), (\ref{BPS8}) and (\ref{CPS8}) to $A_{PS4}$, $B_{PS4}$ and $C_{PS4}$ in   (\ref{dressedtreeA}),
 (\ref{dressedtreeB}) and (\ref{dressedtreeC}) and expanding the exponent up to  ${\cal O}(\ell_s^{10})$.

\begin{figure}[t!!]
	\centering
	\includegraphics[width=3in]{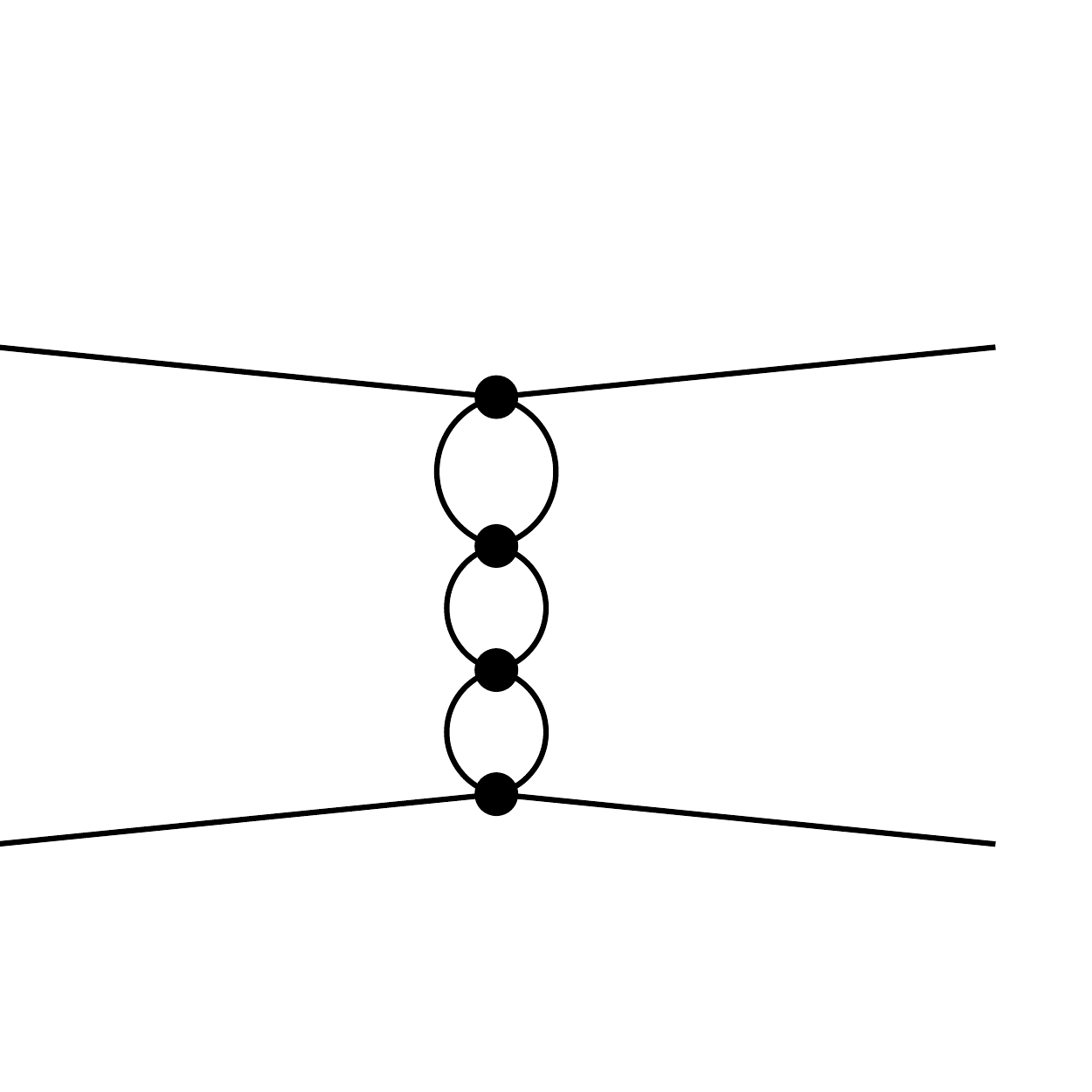}
	\caption{Three loop quartic large D diagram.}
		\label{fig:3loops}
\end{figure}

 For the first time in our calculations these amplitudes exhibit non-local real parts. As a consequence the relation between annihilation $A$ and reflection $C$ amplitudes is now more subtle than $A=-C$, as
 in (\ref{dressedtreeA}),  (\ref{dressedtreeC}). In particular only the annihilation amplitude acquires an imaginary part in the physical region. However, it is straightforward to check that the crossing relation
 (\ref{crossingA}) holds, as it should be.

The dressing procedure allowed us to obtain a ${\cal O}(\ell_s^{10})$ Poincar\'e invariant amplitude corresponding to a particular choice of counterterms (which at this order include $S_1$, $S_2$ and additional counterterms at 
${\cal O}(\ell_s^{8})$ and ${\cal O}(\ell_s^{10})$). In principle, it is a matter of straightforward one loop and tree level calculations including these counterterms to extend the dressing arguments above and to obtain an expression
for the most general  ${\cal O}(\ell_s^{10})$ Poincar\'e invariant amplitude. We will not provide it here, but it is important to stress
 that these counterterms cannot change the ${\cal O}(\ell_s^8)$ logarithmic terms, so these are universal. 
 
 These logarithmic terms disappear at $D=4$, which is related to the existence of integrable non-Poincar\'e invariant theories with annihilations constructed in \cite{Cooper:2014noa}, similarly to how the absence of logarithmic terms at one loop is related to the existence of the GGRT theory. 
We also checked that at three loops the $D^3$ part of  dressed amplitudes  agrees with the brute force result including the leading  large $D$ diagrams, which include a three loop bubble diagram, as shown in Fig.~\ref{fig:3loops}, and also two and one loop diagrams with evanescent  counterterms $S_{E1}$, $S_{E2}$.
 \section{Future Directions}
\label{sec:last}
\begin{figure}[t!!]
	\centering
	\includegraphics[width=2.5in]{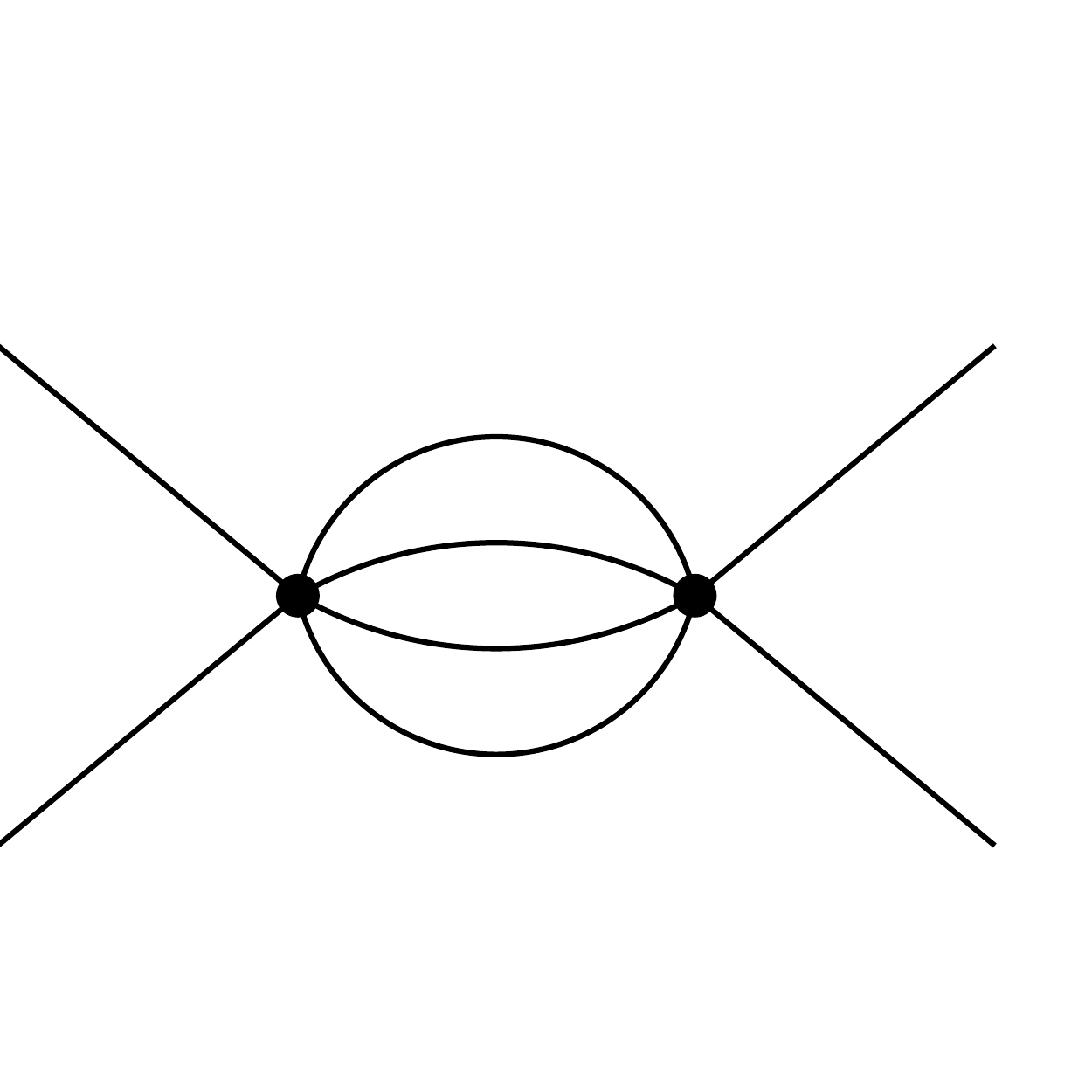}
	\caption{Three loop sextic diagram.}
		\label{fig:2sextics}
\end{figure}
It is quite surprising that in the Nambu--Goto theory relatively straightforward  arguments allow to arrive at what otherwise would be the result of a tedious four loop calculation. Furthermore, the analytical structure of the result is surprisingly simple, albeit  the scattering is non-trivial. The main idea behind our arguments is to make use of the existence  of certain (semi)integrable theories, such as the GGRT theory, and the dressed Polchinski--Strominger.  In the future we plan to push these efforts further. Ideally, one may hope 
that these methods may be promoted into a systematic way of building up perturbative expansions around the GGRT theory.

The results obtained in the current paper raise a number of questions. For instance, it is interesting to understand in more details the structure of the undressed theory $S_{PS}$.
All its $h=2$ vertices are uniquely fixed, and can be determined by inspecting multiparticle one loop amplitudes in the Nambu--Goto theory. It appears feasible to calculate their explicit form building up on the current algebra techniques of \cite{Cooper:2014noa,Dubovsky:2015zey}. At higher weights it is tempting to speculate that it is consistent to set to zero all operators which do not get generated starting from the $h=2$ part of $S_{PS}$.
If true, this implies that there is a large family of operators in the Nambu--Goto theory which do not exhibit logarithmic running, extending the two loop operators $S_1$, $S_2$.

As far as a calculation of the two-to-two scattering amplitudes goes, in this paper we reached 
a natural threshold. Starting from five loops two particle unitarity is violated and one may expect the analytical structure of the amplitude to become more complicated. 
Using the (un)dressing formalism, calculation of this amplitude requires evaluating of two-loop diagrams with three $S_{PS4}$ vertices and three loop diagrams
with two $S_{PS6}$ vertices (see Figure~\ref{fig:2sextics}).
 We plan to address these and other related questions in the near future.

\section*{Acknowledgements}
We are grateful to Raphael Flauger, Victor Gorbenko and Mehrdad Mirbabayi for numerous fruitful discussions. This work was supported in part by the NSF CAREER award PHY-1352119.

\bibliographystyle{utphys}
\bibliography{EffectiveString}
\end{document}